\documentclass{sig-alt-full}

\usepackage{latexsym}
\usepackage{gastex}

\usepackage{tikz}
\usepackage[amsfonts]{logic-thm}

\newtheorem{theorem}{Theorem}
\newtheorem{proposition}[theorem]{Proposition}

\newtheorem{remark}[theorem]{Remark}

\def\arr#1{\stackrel{#1}{\longrightarrow}}

\def\Dd{{\cal D}}
\def\Tt{{\cal T}}
\def\Rr{{\cal R}}
 \def\Xx{{\cal X}}
\def\sat{{\models}}
\def\S{{\Sigma}}
 
\def\Qq{{\cal Q}}
\def\Ff{{\cal F}}

\def\root{\text{root}}
\def\ra{\rightarrow}
\def\rsa{\rightsquigarrow}
\def\et{\tau}
\def\Xx{{\mathcal{X}}}
\def\cond{\text{\emph{cond}}}
\def\self{\text{\emph{self}}}
\def\locator{\text{\emph{locator}}}

\def\del{\text{\emph{del}}}
\def\new{\text{\emph{append}}}

\def\ren{\text{\emph{ren}}}

\def\app{\text{\emph{append}}}

\def\post{\text{\emph{post}}}

\def\body{\text{\emph{body}}}
\def\head{\text{\emph{head}}}

\def\pred{\text{\emph{Pred}}}
\def\succ{\text{\emph{Succ}}}

\def\Init{\text{\emph{Init}}}

\def\init{\text{\emph{Init}}}
\def\nexptime{\textsf{NexpTime}}

\begin{document}

\title{Verifying Recursive Active Documents \\ with Positive Data Tree Rewriting}

\numberofauthors{3}

\author{
\alignauthor
Blaise Genest\\
      \affaddr{CNRS, IRISA, Rennes, France}\\
\alignauthor
Anca Muscholl\\
      \affaddr{LaBRI, Universit\'e Bordeaux 1, France}\\
\alignauthor Zhilin Wu\\
      \affaddr{LaBRI, Universit\'e Bordeaux 1, France}\\
}

\date{November 2009}

\maketitle

\begin{abstract}
  This paper proposes a data tree-rewriting framework for modeling
  evolving documents. The framework is close to Guarded Active XML, a platform
  used for handling XML repositories evolving through web services.
  We focus on automatic verification of properties of
  evolving documents that can contain data from an infinite domain.
  We establish the boundaries of decidability,
  and show
  that verification of  a {\em positive} fragment that can handle recursive service calls is decidable.
We also consider bounded model-checking in our data tree-rewriting
  framework and show that it is $\nexptime$-complete.
\end{abstract}

\keywords{Active documents, Guarded Active XML, automated
  verification, data tree rewriting, tree decompositions.}

\section{Introduction}

From static in house solutions, databases have become more and more
open to the world, offering e.g.~half-open access through web
services.  As usual for open systems, their design requires a careful
static analysis process, helping to guarantee that no malicious client
may take advantage of the system in a way for which the system was not
designed.  Static analysis of such systems very recently brought
together two areas - databases, with emphasis on semi-structured XML data, and
automated verification, with emphasis on model-checking infinite-state
systems.  Systems modeling dynamical evolution of data are pretty
challenging for automated verification, as they involve feedback loops
between
semi-structured data, possibly with values from
unbounded domains, and the workflow of services. If each of these
topics has been studied extensively on its own, very few papers tackle
decidability of algorithms when
all aspects are present at the
same time.

An interesting platform emerged recently for using XML repositories
evolving through web services, namely Active XML (AXML)
\cite{abm08}. These are XML-based documents that evolve dynamically,
containing implicit data in form of embedded service calls. Services
may be recursive, so the evolution of such documents is both
non-deterministic and unbounded in time. A first paper analyzing the
evolution of AXML documents considered \emph{monotonous} documents
\cite{abm04}.  With this restriction, as soon as a service is enabled
in an AXML tree $T$, then from this point on the service cannot be
disabled, and calling it  can only extend $T$. In particular, information
cannot be deleted and subsequent service calls return answers that
extend previous answers. Recently, a workflow-oriented version of
AXML was proposed in \cite{ASV08}: the \emph{Guarded AXML} model (GAXML
for short) adds
guards to service calls, thus
controlling the possible evolution of active documents.  Decidability
in co-2$\nexptime$ of static analysis for {\em recursion-free GAXML}
w.r.t~Tree-LTL properties was established in \cite{ASV08}. The
crucial restriction needed for decidability there is a uniform bound
on the number of possible service calls. Compared to \cite{abm04},
service invocation can terminate, and more importantly, negative guards can
be used. Even more importantly, verification tasks are more complex
and challenging because of the presence of unbounded data. However,
the model relies on a rigid semantics of what a service call can do,
and how (using a workspace etc).
For instance, deletion of data is
not possible.

In this work, our aim is twofold. First, we aim at extending the GAXML
model in a uniform way, by expressing the effect of embedded service
calls in form of tree rewriting rules. Our model DTPRS (\emph{data
  tree pattern rewriting systems}) is based on the same
basic ingredients as GAXML, which are tree patterns for guards and
queries. However, our formalism allows a user to describe several
possible effects of a service call: materialization of implicit data
like in (G)AXML, but also deletion and modification of
existing document parts. This model is a simplified version of the
TPRS model proposed in \cite{GMSZ08}, but it can additionally handle data from infinite domains.

Our second, and main objective is to get decidability of static
analysis of DTPRS without relying on a bound on the number of service
calls. For doing that, we use a technique that emerged in the
verification of particular infinite-state systems, such as Petri nets
and lossy channel systems. The main concept is known in verification
as \emph{well-structured transition systems} (WSTS for short)
\cite{acj96,FS01}. WSTS are one example of infinite-state systems
where (potentially) infinite sets of states can be represented (and
effectively manipulated) symbolically in a finite way.
In contrast, \cite{ASV08} uses a small model property which implies an enumeration of trees up to some bound.

Our basic objects are data trees, i.e., trees with labels from an
infinite domain. We view data trees as graphs, and define in a natural
way a well-quasi-order on such graphs. Then we show that a uniform
bound on the length of simple paths in such graphs, together with
positive guards, makes DTPRS well-structured systems
\cite{acj96,FS01}. As a technical tool we use here tree
decompositions of graphs. In a nutshell we trade here recursion against
positiveness, since considering both leads to undecidable static
analysis. We show that for \emph{positive} DTPRS, termination and tree
pattern reachability are both decidable.
Furthermore, we show
that \emph{bounded} model-checking of (\emph{not necessarily positive})
DTPRS is $\nexptime$-complete.
On the other hand, we show that the verification of simple but non positive temporal properties is undecidable even for \emph{positive} DTPRS.



\medskip

\emph{Related work:}
Verification of web services often ignores unbounded data (c.f.
e.g.~\cite{hbcs03,fbs04}).  On the other hand, several data-driven workflow process models have been proposed.
Document-driven workflow was proposed in \cite{WK05}.
Artifact-based workflow was outlined in \cite{HULL08}, in
which artifacts are used to represent key business entities,
including both their data and life cycles.
An early line of results involving data establishes decidability boundaries for the verification of temporal
(first-order based) properties of a data-driven workflow processes,
based on a relational data model \cite{dsvz06,dsv07,dv08}. This
approach has been recently extended to the artifact-based model
\cite{dhpv09}.


On the verification side,
there is a rich literature on the verification of
well-structured infinite transition systems \cite{acj96,FS01}, ranging
from faulty communication systems \cite{bmosw08} to programs
manipulating dynamic data \cite{abchr08} (citing only a few
recent contributions).
The latter work is one of the few examples where
well-quasi-order on graphs are used.

\emph{Organization of the paper:} In the next section, we fix some
definitions and notations, define the DTPRS model, and illustrate
how to reduce GAXML to our DTPRS model. Then in Section 3, we describe an
example to illustrate the expressivity of DTPRS. In Section 4, we
show that DTPRS with recursive DTD or negated tree patterns are
undecidable. In Section 5 we define positive
DTPRS and prove our decidability results. On the other hand, we show
the undecidability of the verification of general, non-positive temporal
properties in Section 6. Finally in Section 7, we
consider bounded model-checking of (not necessarily positive) DTPRS
and show that the bounded model-checking problem is
$\nexptime$-complete. Omitted proofs can be found in the appendix.

\section{Definitions and notations}

In this paper, documents correspond to labeled,  unranked and
unordered trees.
Fix a finite alphabet $\Sigma$ (with symbols $a,b,c,\ldots$, called
\emph{tags}) and an infinite data domain $\Dd$.  A \emph{data tree}
is a (rooted) tree $T$ with nodes labeled by $\S \cup \Dd$.  A data
tree $T$ can be represented as a tuple $T=(V,E,\root,\ell)$, with
labeling function $\ell : V \ra \S \cup \Dd$. Inner nodes are
$\S$-labeled, whereas leaves are $(\S \cup \Dd)$-labeled. We fix a
finite set of variables $\Xx$ (with symbols $X,Y,Z,\ldots$) that
will take values in $\Dd$, and use $*$ as special symbol standing
for any tag. Let $\Tt$ denote $\S \cup \Xx \cup \{*\}$.

A \emph{data constraint}  is a Boolean combination of
relations $X = Y$, with\footnote{For simplicity  we disallow here
explicit data constants $X=d$ ($d \in \Dd$): they can be
simulated by tags from $\Sigma$.} $X,Y \in \Xx$.

A \emph{data tree pattern} (DTP) $P=(V,E,\root,\ell,\et, \cond)$ is
a (rooted) $\Tt$-labeled tree, together with an edge-labeling
function $\et : E \ra \{|,||\}$ and a data constraint $\cond$. As
usual, $|$-labeled edges denote child edges, and $||$-labeled edges
denote descendant edges. Internal nodes are labeled by $\S \cup
\{\ast\}$, and leaves by $\Tt$.
A \emph{matching} of
a DTP $P$ into
a data tree $T$ is defined as a mapping preserving the root, the
$\S$- and $\Dd$-labels (with $\ast$ as wildcard), the child- and the
descendant relations, satisfying $\cond$ and mapping $\Xx$-labeled
nodes to $\Dd$-labeled ones.  In particular, a relation $X=Y$ ($X,Y
\in \Xx$) means that the corresponding leaves are mapped to leaves
of $T$ carrying the same data value. If the mapping above is
injective, then it is called an \emph{injective} matching of $P$
into $T$.

A \emph{relative} DTP is a DTP with one designated
node \self. A relative DTP $(P,\self)$ is matched to a pair $(T,v)$,
where $T$ is a tree and $v$ is a node of $T$.

We consider \emph{Boolean combinations} of (relative) DTPs.
The patterns therein are matched \emph{independently} of each other
(except that nodes designated by \self\ must be matched to
the same node of $T$), and the
Boolean operators are interpreted with the standard meaning.

A \emph{data tree pattern query} (DTPQ) is of the form $\body \rsa
\head$, with $\body$ a DTP and $\head$ a tree such that
\begin{itemize}
\item the internal nodes of $\head$ are labeled by $\S$ and its leaves are
labeled by $(\S \cup \Dd \cup \Xx)$,

\item every variable occurring in $\head$ also occurs in $\body$,

\item there is at least one variable occurring in $\head$, i.e., at
 least one leaf of $\head$ is labeled by $\Xx$.
\end{itemize}

Let $T$ be a data tree and $Q=\body \rsa \head$ be a DTPQ. The
evaluation result of $Q$ over $T$ is the forest $Q(T)$  of all
instantiations of $\head$ by matchings $\mu$ from $\body$ to $T$. A
\emph{relative} DTPQ is like a DTPQ, except that its $\body$ is
a relative pattern.

A \emph{locator} is a relative DTP $L$ with additional labels from
the set $\{\new,\del\} \cup \{ren_a \mid a \in \S\}$. The labels
$\new$ and $\ren_a$ are not exclusive and can be attached only to nodes of
$L$ that are labeled by a tag (that is by $\S \cup \{\ast\}$ but not
by $\Dd \cup \Xx$).
Nodes not labeled by $\new,\ren_a$ can be labeled by $\del$
(even data nodes), such
that the descendants of a node labeled by $\del$ are labeled by
$\del$, too. The intuition behind this definition is to provide a
context for data tree rewriting rules, together with some possible actions on
this context: renaming, deletion, appending.

A \emph{data tree pattern rewriting rule} (DTP rule) $R$ is a tuple
$(L,G,\Qq,\Ff,\chi)$ with:
\begin{itemize}
\item $L$ is a \locator,
\item $G$  is a Boolean combination of
  (relative) DTPs (the \emph{guard} of $R$),
\item $\Qq$ is a finite set of relative DTPQs,
\item $\Ff$ is a finite set of forests with internal
 nodes labeled by $\S$ and leaves labeled by $\S \cup \Dd \cup \Xx
 \cup \Qq$,

\item $\chi$ is a mapping from the set of nodes of $L$ labeled by
 \new\ to $\Ff$.
\end{itemize}


\medskip

A \emph{DTP rewriting system (DTPRS)}  is a pair $(\Rr,\Delta)$
consisting of a set $\Rr$ of DTP rules and  a \emph{static
invariant} $\Delta$, consisting of a \emph{DTD} and a \emph{data
invariant}, i.e.~a
Boolean combination of DTPs. We assume that the static invariant
$\Delta$ is preserved by the rewriting rules $\Rr$. As usual for
unordered trees, a DTD is defined as a tuple $(\Sigma_r, \Pp)$ such
that $\Sigma_r$ is the set of allowed root labels, and $\Pp$ is a
finite set of rules $a \ra \psi$ such that $a \in \Sigma$ and $\psi$
is a Boolean combination of inequalities of the form $|b| \ge k$,
where $b \in \Sigma \cup \{dom\}$ ($dom$ is a symbol standing for
any data value), and $k$ is a non-negative integer. A \emph{positive
  DTD} is one where the Boolean combinations above are positive.

We now define formally the semantics of a transition by DTP rules.
So let $T=(V,E,\root,\ell)$ be a data tree  (with $T \sat \Delta$) and let
$R=(L,G,\Qq,\Ff,\chi)$ be a DTP rule.

\begin{itemize}

\item Let $\mu$ be an {\em injective} matching from $L$ to $T$. Let
  $\nu$ be the assignment of data values to variables in $L$ such that
  $\nu(X)=\ell(\mu(v))$ for every $v$ labeled by $X \in \Xx$ in $L$.

\item 
For each variable $X \in \Xx$ we denote its
  evaluation as $X(T)$, with $X(T)=\nu(X)$ if defined, and
    $X(T)$ \textbf{a fresh data value otherwise}.
Here a fresh data value is a data value which
 does not appear anywhere else in $T$. Furthermore, it is required that all the \emph{new} variables of $R$,
i.e.~variables occurring in $\Ff$, but not in $L$, should take \textbf{mutually distinct} fresh values. For each forest $F \in
  \Ff$, we denote its evaluation by $F(T)$, by replacing labels $Q \in
  \Qq$ by $Q(T)$ and labels $X \in \Xx$ by $X(T)$. Recall that all
  queries $Q \in \Qq$ are evaluated relatively to $\mu(\self)$.

\item A data tree $T'$ can be obtained from $T$ by
\begin{itemize}
\item deleting subtrees rooted at nodes $\mu(v)$ whenever $v$ is
  labeled by $\del$ in $L$,

\item changing the tag of a node $\mu(v)$ to $a$
 whenever  $v$ is labeled by $\ren_a$ in $L$,

\item appending $F(T)$ as a subforest of nodes $\mu(v)$ whenever
  $v$ is labeled by $\new$ in $L$ and $\chi(v)=F$,

\item every other node of $T$ keeps its tag or data.
\end{itemize}

\item The rule $R$ is enabled on data tree $T$ if there exists an
  \emph{injective} matching $\mu$ of $L$ into $T$ such that (1) $G$ is
  true on $(T,\mu(v))$ with $v$ labeled by \self\ in $L$, and (2)
  there is a data tree $T'$, obtained from $T$ and $\mu$ by the
  operations specified above, satisfying $T' \sat \Delta$.
\end{itemize}

Let $T \arr{R} T'$ denote the transition from $T$ to $T'$ using DTP
rule $R \in \Rr$.

\begin{remark}

 \begin{enumerate}

\item The injectivity of the matching  $\mu$ ensures that the
  outcome $T'$ is
  well-defined. In particular, no two nodes with label $\del$ and
  $\new$ (or $\ren_a$), resp., can be mapped to the same node in the
  data tree. However, mappings used for guards or queries
  are - as usual -  non injective.

\item
For the new variables occurring in $\Ff$, but not in $L$, we
 choose mutually distinct fresh values. We could have chosen
 arbitrary values instead, and enforce the fact that they are fresh and mutually distinct
 \emph{a posteriori} using the invariant $\Delta$.
 In this case, the invariant needs negation.
 The invariant (or the locator) can be also used to enforce that the
 (arbitrarily)  chosen values already occur in $T$.
 This kind of  invariant would be positive.

\item In our definition of DTP rules, it might appear that guards are redundant
  wrt.~the locator. But notice
  that this only concerns positive guards.
 \end{enumerate}
\end{remark}

Given a DTPRS $(\Rr,\Delta)$, let $T \arr{} T^\prime$ denote the
union of $T \arr{R} T^\prime$ for $R \in \Rr$, and $T \arr{+}
T^\prime$ (or $T \arr{\ast} T^\prime$) denote the transitive (or
reflexive and transitive) closure of $T \arr{} T^\prime$. Moreover,
let $\Tt^*_\Rr(T)$ denote the set of trees that can be reached from
a data tree $T$ by rewriting with DTP rules from $\Rr$, i.e.
$\Tt^\ast_\Rr(T) = \{T^\prime \mid T \arr{\ast} T^\prime\}$. For a
set of data trees $\cal{I}$, let $\Tt^*_\Rr(\cal{I})$ be the union of
$\Tt^*_\Rr(T)$, for $T \in \cal{I}$.

We are interested in the following questions, given a DTPRS
$(\Rr,\Delta)$:
\begin{itemize}
\item \emph{Pattern reachability}: Given a  DTP $P$ and a
  set of initial trees\footnote{We require that every tree in $\Init$
    satisfies $\Delta$.} $\Init$, given as the conjunction of a DTD
  and a Boolean
combination of DTPs, is there some
  $T \in \Tt^*_\Rr(\Init)$ such that $P$ matches $T$?

\item \emph{Termination:} Given an initial data tree $T_0$, are
  all rewriting paths $T_0
  \ra T_1 \ra \cdots$ starting from $T_0$ finite?
\end{itemize}

The reason for the fact that termination of DTPRS is defined above
w.r.t~a single initial data tree is that termination from a set of
initial trees is already undecidable without data (see Section
\ref{s:undec}).

%

\subsection{Reduction from GAXML to DTPRS}

DTPRS is a quite powerful model, which allows to model e.g.~Guarded
Active XML (GAXML)
\cite{ASV08}. We show this translation on two main GAXML steps: call
and return of services. For completeness, we briefly recall the main
features of GAXML here. AXML trees contain embedded function calls
that are evaluated (via tree pattern queries) in a workspace. GAXML
adds call and return guards, that control the function call and the
return of the result (as sibling of the call node). Functions can be
internal or external. The external ones return some arbitrary forest
that is consistent with the static invariant $\Delta$.

We describe here how to model an (internal) function call in GAXML
with DTPRS. Let $f \in \Sigma$ be a function associated with the
argument query $Q$ and the call guard $G$.  The associated DTP rule
has the same guard $G$, the set of queries $\Qq = \{Q\}$, and $\Ff =
\{T_f,T_X\}$ is the set defined below:
\begin{itemize}
\item $T_f$ is a tree with three nodes, the root being labeled by $f$,
  and the two leaves labeled by the query $Q$ and by the variable $X$,
  respectively.
\item $T_X$ is a tree with a unique node labeled by the variable $X$,
\end{itemize}

\begin{figure}[ht]
\begin{center}
\scalebox{0.8}{
\setlength{\unitlength}{2mm}
\begin{picture}(20,10)(0,3)

\gasset{Nw=4,Nh=2,Nmr=1}

\node(v1)(10,10){root}

\node(v2)(0,0){$!f$}

\node(v3)(15,5){WS}

\gasset{Nframe=n,Nw=4,Nh=2,Nmr=0}

\node(v2-s)(-.5,2){\self}

\node(v2-a)(9,0){$\ren_{?f}$,$\app(T_X)$}

\node(v3-a)(22,5){$\app(T_f)$}

\node(vv1)(10.5,10){}

\node(vv2)(0.5,0){}

\drawedge[AHLength=0,AHlength=0](v1,v2){}
\drawedge[AHLength=0,AHlength=0](vv1,vv2){}
\drawedge[AHLength=0,AHlength=0](v1,v3){}

\end{picture}
}
\end{center}
\caption{The locator for service invocation.}
\label{f.invocation}
\end{figure}
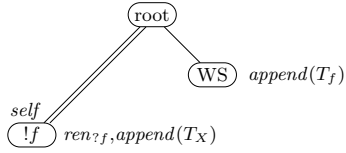

The locator $L$ is given in Figure~\ref{f.invocation}. Finally, $\chi$
maps the $!f$-node  to $T_X$ and the WS-node to $T_f$. Applying
the DTP rule amounts in evaluating $Q$ to get the arguments of the
call, writing them into the workspace WS, and creating a fresh
identifier $X$ that it copied both below WS  and below the node with
the function call  (aka return address for $f$, see below).

We describe now how to model the return of an (internal continuous)
function $f \in \Sigma$ associated with the return query $Q$ and the
return guard $G$.  The associated DTP rule has the same guard $G$,
the set of queries $\Qq = \{Q\}$, $\Ff = \{T_Q\}$ has a unique tree
with a unique node labeled by the query $Q$, and the locator is
given in Figure~\ref{f.return} with the data constraint $X=Y$:

\begin{figure}[ht]
\begin{center}
\scalebox{0.8}{
\setlength{\unitlength}{2mm}
\begin{picture}(20,15)(0,-4)

\gasset{Nw=4,Nh=2,Nmr=1}

\node(v1)(10,11){root}
\node(v12)(0,5){$\ast$}
\node(v2)(0,0){$?f$}
\node(v3)(18,5){WS}
\node(v4)(18,0){$f$}
\node(v5)(18,-5){$Y$}
\node(v6)(0,-5){$X$}

\gasset{Nframe=n,Nw=4,Nh=2,Nmr=0}

\node(v2-s)(-1.5,2){\self}

\node(v12-a)(6.5,5){$\app(T_Q)$}

\node(v2-a)(4.5,0){$\ren_{!f}$}

\node(v4-a)(21.5,0){$\del$}

\node(v5-a)(21.5,-5){$\del$}

\node(v6-a)(3.5,-5){$\del$}

\node(vv1)(10.5,11){}
\node(vv2)(0.23,4.7){}

\drawedge[AHLength=0,AHlength=0](v1,v12){}
\drawedge[AHLength=0,AHlength=0](vv1,vv2){}
\drawedge[AHLength=0,AHlength=0](v12,v2){}
\drawedge[AHLength=0,AHlength=0](v1,v3){}
\drawedge[AHLength=0,AHlength=0](v3,v4){}
\drawedge[AHLength=0,AHlength=0](v4,v5){}
\drawedge[AHLength=0,AHlength=0](v2,v6){}

\end{picture}
}
\caption{The locator for service return.}
\label{f.return}
\end{center}
\end{figure}
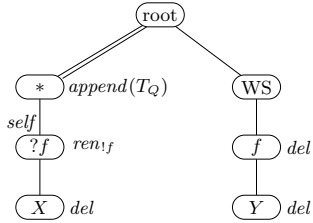
Finally, $\chi$ maps the node labeled $\app$ with $T_Q$.  This DTP
rule locate where the call was computed in the workspace WS using
$X$, evaluates $Q$ to get the return of the call, puts the result of
the return query as a sibling of $?f$, and deletes the associated data in the
workspace, as well as the identifier $X$.

\section{MailOrder example}\label{s:mo-example}

\def\play{\textit{Play.com}}
\def\ccat{\text{CCatalog}}
\def\pcat{\text{PCatalog}}

\def\cid{\text{CId}}
\def\name{\text{Name}}
\def\eml{\text{Email}}

\def\prdt{\text{Product}}
\def\pid{\text{PId}}
\def\price{\text{Price}}

\def\cust{\text{Customer}}
\def\cart{\text{Cart}}
\def\itm{\text{Item}}
\def\ord{\text{Order}}
\def\oid{\text{OId}}
\def\sta{\text{State}}
\def\ordered{\text{payment}}
\def\process{\text{Processed}}
\def\post{\text{Posted}}
\def\bill{\text{Bill}}
\def\pay{\text{Paid}}
\def\recv{\text{Received}}
\def\payback{\text{PaidBack}}

\def\selec{\text{select}}

The following is a DTPRS description of the basic functionalities of a
MailOrder system for \emph{Play.com}. For simplicity, we represent
only what happens on the \emph{Play.com} peer, although we could also
model client peers, bank peers etc.

The \emph{Play.com} example can be compared with the \emph{MailOrder} example
in \cite{ASV08}.
Syntactically, GAXML uses guards and queries.  Most of the
time, guards and queries are very simple and can be encoded in
the locator of DTP rules. In this case, we omit the {\em self} label
in our rules.
Unlike \emph{MailOrder}  we can express
deletion with DTPRS, and thus model the selection of the products in the
cart (adding and deleting products), and also handle an inventory (how
many items of a product remain - each time an item  is added to a cart,
it is also deleted from the inventory). More importantly, compared
with the recursion-free decidable restriction of GAXML, we are able to
represent the  process of many customers ordering many different
products in our decidable fragment.

On  the \emph{Play.com} peer, there are a product catalog, a
customer catalog, a set of carts and a set of orders.  The inventory
is encoded in the product catalog: if there are three items of a
product in the inventory, then there are three tokens as children of
the product. Each cart is associated with a customer (at first
anonymous, and he can later login in as a registered member).  The
cart is first in the select mode, which allows the associated
customer to add/delete products. Then the customer can check out,
the cart gathers the different prices for the products into a bill,
and goes into payment mode. When the customer pays, a corresponding
order is created with a receipt, the customer is disconnected and
the cart is deleted.

\begin{figure}[ht]
\begin{center}
\scalebox{0.78}{ \setlength{\unitlength}{2mm}
\begin{picture}(20,22)(15,-4)

\gasset{Nadjust=w,Nw=4,Nh=2,Nmr=1}

\node(v1)(24,15){\play}

\node(v2)(5,10){\ccat}

\node(v4)(24,10){\cart}

\node(v5)(44,10){\pcat}

\node(v6)(5,5){\cust}

\node(v9)(16.5,5){\text{log}}

\node(v99)(16.5,0){\cid}

\node(v12)(32.5,5){\selec}


\node(v13)(44,5){\prdt}

\node(v14)(0,0){\cid}

\node(v15)(5,0){\name}

\node(v16)(11,0){\eml}

\node(v18)(24,5){\text{products}}

\node(v88)(26.5,0){\pid}

\node(v188)(21,0){\pid}

\node(v21)(37.5,0){\pid}

\node(v22)(42,0){\name}

\node(v23)(47,0){\price}

\node(v24)(52,0){\text{token}}

\drawedge[AHLength=0,AHlength=0](v1,v2){}
\drawedge[AHLength=0,AHlength=0](v1,v4){}
\drawedge[AHLength=0,AHlength=0](v1,v5){}

\drawedge[AHLength=0,AHlength=0](v9,v99){}

\drawedge[AHLength=0,AHlength=0](v18,v88){}
\drawedge[AHLength=0,AHlength=0](v18,v188){}

\drawedge[AHLength=0,AHlength=0](v2,v6){}


\drawedge[AHLength=0,AHlength=0](v4,v9){}
\drawedge[AHLength=0,AHlength=0](v4,v18){}
\drawedge[AHLength=0,AHlength=0](v4,v12){}

\drawedge[AHLength=0,AHlength=0](v5,v13){}

\drawedge[AHLength=0,AHlength=0](v6,v14){}
\drawedge[AHLength=0,AHlength=0](v6,v15){}
\drawedge[AHLength=0,AHlength=0](v6,v16){}


\drawedge[AHLength=0,AHlength=0](v4,v18){}

\drawedge[AHLength=0,AHlength=0](v13,v21){}
\drawedge[AHLength=0,AHlength=0](v13,v22){}
\drawedge[AHLength=0,AHlength=0](v13,v23){}
\drawedge[AHLength=0,AHlength=0](v13,v24){}


\drawedge[AHLength=0,AHlength=0,curvedepth=4,dash={1.5}0](v99,v14){}
\drawedge[AHLength=0,AHlength=0,curvedepth=-4,dash={1.5}0](v188,v21){}
\drawedge[AHLength=0,AHlength=0,curvedepth=3,dash={1.5}0](v21,v88){}

\node[fillcolor=white,Nmr=0](v30)(8,-4){\text{Blaise}}

\node[fillcolor=white,Nmr=0](v31)(33.5,-3.3){\text{9221}}

\end{picture}
} \caption{A configuration of the \play\ system, where the
registered customer Blaise has twice the same product in his cart,
with PID 9221. There is one additional product with PID 9221 left. }
\label{fig-play-com-dtd}
\end{center}
\end{figure}
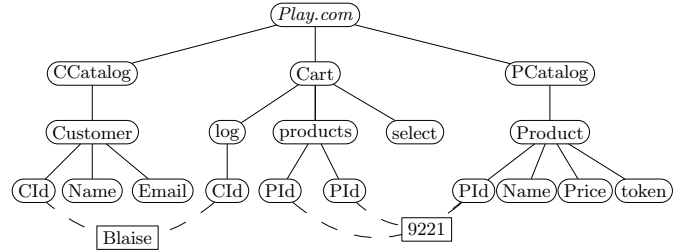

A simple example of a configuration of \emph{Play.com} is illustrated
in Figure~\ref{fig-play-com-dtd}. We represented a data value only
when it is used by at least two different nodes.
%
%
%
%
%
%
%
%

Some key rewriting rules are described in the following. We only
describe nontrivial components in these rules. Tree patterns are
represented below in term form, with descendant edges preceded by a $-$
symbol.

\begin{itemize}

\item An anonymous customer can connect to \play\ with the rule
\emph{create-cart}.
\begin{itemize}
\item $L:=[\play_{\app(F)}]$
\item $F:=[\cart]([\text{nolog}]([\cid](X)),[\text{products}],[\selec] )$.
Notice that here $X$ is a fresh data value.
\end{itemize}

\item An anonymous customer can login as a registered member with the
  rule \emph{login}. As the customer peer is not modeled, we do not
  handle the check of a password, although it would not be a problem
  to do so in our framework.

\begin{itemize}
\item $L:=[\play](-[\cust]([\cid](X))$,
\\ $[\cart_{\app(F)}]([\text{nolog}_{\del}]))$
\item $F:=[\text{log}]([\cid](X))$.
Here, $X$ is the same \pid \, for the cart and for the customer.
\end{itemize}

\item The rule \emph{Add-Product} adds a new product into the cart
  (and deletes a token from the inventory).
\begin{itemize}
\item $L:=[\play]([\cart]([\text{products}_{\app(F)}]),$\\
$-[\prdt]([\pid](X),[\text{token}_{\del}]))$
\item $F:=[\pid](X)$.
\end{itemize}

\item The rule \emph{Delete-Product} deletes a product from the cart
  (and puts the token back).
\begin{itemize}
\item $L:=[\play]([\cart]([\selec],-[\pid_{\del}](X)),$\\
$-[\prdt_{\app(F)}]([\pid](X)))$
\item $F:=[\text{token}]$.
\end{itemize}

\item The rule \emph{Check-out} checks whether the cart is nonempty
  and retrieves the prices of products in the cart into a bill through a
  query. It changes the mode of the cart from {\em select} to
  {\em payment}.
\begin{itemize}
\item $L:=[\play]([\cart_{\self,\app(F)}](-[\pid],$
\\ $[\selec_{\ren_{\ordered}}]) )$
\item $F:=[\text{Bill}](Q)$, and $Q:=\body \rsa Y$ is the query with $\body =
[\play]([\cart_{\self}] (-[\pid](X)),\\
-[\text{Product}]([\pid](X),[\text{Price}](Y)) )$.
\end{itemize}

\item The customer can pay with the rule \emph{Pay}, and a
  corresponding order is created. For simplicity, it
  disconnects the customer, and transforms the cart into an order. The
  order contains the customer ID, an order ID (a fresh unique
  identifier), and the total price (sum of the prices of each
  items). As we model prices by data
  values and we do not use any arithmetics, the total price is a fresh data
  value. The only important thing
  is that this data value {\em Total} is the same as the one registered
  in the bank account, which we could check for equality (although we do not
  explicitly model the bank here). The order does not recall the
  individual PID of products since there will be no more products to
  put back in the inventory (and anyway the product can be later
  removed from the catalog).

\begin{itemize}
\item $L:=[\play]([\cart_{\app_F,\ren_{\ord}}]( [\text{Bill}_{\del}]$,\\
$[\text{products}_{\del}],[\ordered_{\ren_{\text{paid}}}],[\log_{\ren_{\text{cust}}}]))$
\item $F:=[\text{Receipt}]([\oid](X),[\text{Total}](Y))$
\end{itemize}


\item The rule \emph{Add-member} allows a customer to register as a member.
\begin{itemize}
\item $L:=[\play]([\ccat_{\app(F)}])$
\item $F:=[\cust]([\cid](X),[\name](Y),[\eml](Z))$
\end{itemize}
\end{itemize}

We do not specify the following rules here, which are easy to come
up with: {\em shipped}, {\em delivered}, {\em add product to
catalog} etc.

Notice that this example is intentionally not correct.  Indeed, as
there is no check of the state in which the cart is at rule {\em
  Add-Product}, it is possible that a bill is produced after a check
out, and the customer can still add products with the aforementioned
rule which will never be accounted for in the bill. We will show
later that we can decide whether such a problem occurs or not. To
fix this problem, it suffices to check in the locator $L$ of
\emph{Add-Product} rule that the cart is in the select mode.

\section{Undecidability}\label{s:undec}

As one might expect, analysis of DTPRS is quickly undecidable --
and sometimes already without using any unbounded data. The proof of the
proposition below is obtained by straightforward simulations of
2-counter machines.

\begin{proposition}\label{p:recneg}
  Both termination and pattern reachability for DTPRS $(\Rr,\Delta)$
  are undecidable whenever one of the following holds:
  \begin{enumerate}
  \item the DTD in $\Delta$ is recursive,
\item either guards in $\Rr$ or the invariant $\Delta$ contain negated DTPs.
  \end{enumerate}
The above result holds even without data.
\end{proposition}




The next result shows that with data, we can relax both conditions
above and still get undecidability of DTPRS. The main idea is to use
data for creating long horizontal paths (although trees are supposed
to be unordered). Such horizontal paths can be obtained e.g.~with a
tree of depth 2, having subtrees of height one with  2 leaves each,
say labeled by data values $d_i,d_{i+1}$. Assuming all $d_i$ are
distinct (and distinguishing $d_1$) we get a linear order on these
subtrees.

\begin{theorem}\label{t:fresh}
  Both termination and pattern reachability are undecidable for DTPRS
  $(\Rr,\Delta)$ such that (1) the DTD in $\Delta$ is non-recursive
  and (2) all DTPs in guards from $\Rr$ and the invariant $\Delta$ are
  positive.
\end{theorem}

 \begin{proof}
   We reduce Post correspondence problem (PCP) first to pattern reachability. We may assume that our
   PCP instance $(u_i,v_i)_{1 \le i \le n}$ is such that the following
   holds for every non-empty sequence $i_1,\ldots,i_k$ of indices:
   \begin{itemize}
   \item Either $U=u_{i_1} \cdots u_{i_k}$, $V=v_{i_1} \cdots v_{i_k}$
     are incomparable, or $V$ is a prefix of $U$. In the latter case we
     call $(U,V)$ a partial solution.
 \item If  $(U,V)$ and $(Uu_i,Vv_i)$ are  partial solutions and $U
   \not= V$, then
   either $Uu_i=Vv_i$ or $Vv_i$ is a prefix of $U$.
 \item Every solution starts with with the pair $(u_1,v_1)$ and ends
   with $(u_n,v_n)$.
   \end{itemize}
  It is not hard to
 verify that  the usual Turing machine reduction to PCP satisfies
 the restrictions above.

 A partial solution $(U,V)$ with $U=a_1 \cdots a_n$, $V =a_1 \cdots
 a_{m-1}$ will be represented by the data tree below. In this data tree
 the leaves are labeled by data $d_i$, with $d_i \not= d_j$ for all $i
 \not= j$. Moreover, notice that the last position has the special
 marker $\$$, and the first position in $U$ without $V$ has the
 special marker $\#$.

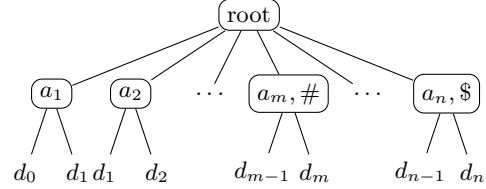
 \begin{tikzpicture}[scale=.7,level 2/.style={sibling distance = 1cm}]
 \node [rounded corners, rectangle, draw] {root}
 child {node[rounded corners, rectangle, draw] {$a_1$}
       child {node {$d_0$}}
       child {node {$d_1$}}
        }
 child  {node[rounded corners, rectangle, draw] {$a_2$}
       child {node {$d_1$}}
       child {node {$d_2$}}
        }
 child {node {$\ldots$}}
 child {node[rounded corners, rectangle, draw] {$a_m,\#$}
       child {node {$d_{m-1}$}}
       child {node {$d_m$}}
        }
 child {node {$\ldots$}}
 child {node[rounded corners, rectangle, draw] {$a_n,\$$}
       child {node {$d_{n-1}$}}
       child {node {$d_n$}}
        };
 \end{tikzpicture}

 With each PCP pair $(u_i,v_i)$, $i<n$, we associate DTPRS rules
 $R_i=(L,G,\Qq,\Ff,\chi)$. For simplicity we describe below the locator
 $L$ and the forest $\Ff$ for $(u_i,v_i)=(aba,bb)$ (the guard $G$ and
 the query set $\Qq$ are
 both empty):

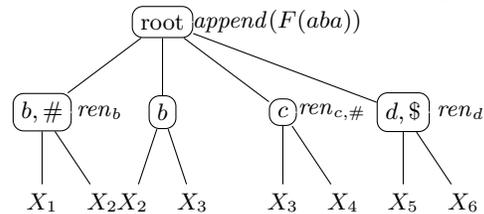
\begin{tikzpicture}[scale=.8,level 1/.style={sibling distance = 2cm},level 2/.style={sibling distance = 1cm}]
  \node [rounded corners, rectangle, draw] {root}
child[grow = right,white,level distance=14ex] {node[black]
{$\app({F(aba)})$}} child {node [rounded corners, rectangle, draw]
{$b,\#$}
      child[grow = right,white,level distance=7ex] {node[black] {$\ren_b$}}
      child {node {$X_1$}}
      child {node {$X_2$}}
      }
child {node [rounded corners, rectangle, draw] {$b$}
      child {node {$X_2$}}
      child {node {$X_3$}}
      }
child {node[rounded corners, rectangle, draw] {$c$}
      child[grow = right,white,level distance=6ex] {node[black] {$\ren_{c,\#}$}}
      child {node {$X_3$}}
      child {node {$X_4$}}
      }
child {node [rounded corners, rectangle, draw]  {$d,\$ $}
      child[grow = right,white,level distance=7ex] {node[black] {$\ren_d$}}
      child {node {$X_5$}}
      child {node {$X_6$}}
      };
\end{tikzpicture}

 We need a rule $R_i$ for each pair of tags $c,d \in\S$ (these are the
 tags at positions $n$ and $m+2$ in the example with $|v_i|=2$).
 The forest $F(aba)$ which will be added under the root node will
 contain 3 trees $T_1,T_2,T_3$, with roots labeled $a$, $b$ and $a,\$$,
 respectively. Tree $T_1$ has two leaves, labeled $X_6$ and
 $X_7$. Tree $T_2$ ($T_3$, resp.) has two leaves,
 labeled $X_7$ and $X_8$ ($X_8$ and $X_9$, resp.). Notice that variable
 $X_6$ occurs in both $L$ and $F(aba)$, whereas $X_7,X_8,X_9$ will take
 fresh (and mutually distinct) values.

 The pair $(u_n,v_n)$ has similar rules, except that we will not append
 any forest to the root, but rename the root with a special marker
 $\surd$. The initial tree $T_0$ is defined as
 expected, from $(u_1,v_1)$. The PCP instance has a solution iff we can
 reach a data tree with root label $\surd$. Notice that all guards and
 the invariant are empty.

 For termination we can modify the above proof in order to ensure that
 executions that do not correspond to partial solutions, are
 infinite. More precisely, if $U,V$ as above is a partial solution, but
 $Uu_i, Vv_i$ is not, then we use a DTP rule associated with
 $(u_i,v_i)$ that forces an infinite execution. In this way,
 termination will hold iff the PCP instance has a solution.
 \end{proof}

We end this section with a remark on the decidability of termination
from an initial set of trees. First we notice that -- already without
data -- DTPRS can simulate so-called \emph{reset Petri nets} \cite{GMSZ08}. These
are Petri nets (or equivalently, multi-counter automata without zero
test) with additional transitions that can reset places (equivalently,
counters) to zero. They can be represented by trees of depth 2, where
nodes at depth one represent places, and their respective number of
children (leaves) is the number of tokens on that place. A DTPRS (without
data) can easily simulate increments, decrements and resets (using
deletion in DTPRS).  It is known that so-called
\emph{structural termination} for reset Petri nets is undecidable
\cite{mayr03}, i.e., the question whether there are infinite
computations from \emph{any} initial configuration, is
undecidable. This implies:

\begin{proposition}
  The following question is undecidable: Given a DTPRS $(\Rr,\Delta)$,
  is there some tree $T_0$ satisfying $\Delta$ and an infinite
  computation $T_0 \arr{} T_1 \arr{} \cdots$ in $(\Rr,\Delta)$? This
  holds already for non-recursive DTD in $\Delta$ and without data constraints
  in DTPs.
\end{proposition}



\section{Positive DTPRS}

In this section we consider  {\em positive DTPRS}, a fragment of DTPRS
for which we show that termination and pattern reachability are decidable.


From Proposition~\ref{p:recneg}, we know that in order to get
decidability, the DTD in the static invariant $\Delta$ must be
non-recursive. For a non-recursive DTD, there is some $B$ such that
every tree satisfying the DTD has depth bounded by $B$. In the
following, we assume the existence of such a bound $B$.  Also from
Proposition~\ref{p:recneg}, we know that for decidability we need to
consider only positive guards and positive data invariants.

However, from Theorem~\ref{t:fresh}, we know that these restrictions
alone do not suffice to achieve decidability. We need to disallow long
linear orders created with the help of data. For this, we introduce
a last restriction, called {\em simple-path bounded}, which is defined in
the following.

Let $T=(V,E,root,\ell)$ be a data tree. \emph{The graph $G(T)$
associated with $T$} is the undirected graph obtained by adding to $V$
the set of data values occurring in $T$, and adding to $E$ the links
between a leaf labeled by a data value and the node representing that value
(see also Figure \ref{fig-play-com-dtd}).
Formally, $G(T)=(V^\prime,E^\prime)$, where $V^\prime=V \cup \{\ell(v) \mid
\ell(v) \in \Dd\}$ and $E^\prime=E \cup \{\{v,d\} \mid \ell(v)=d \in
\Dd\}$. A \emph{simple path} of $T$ is a simple path in $G(T)$, i.e.~a sequence of vertices $v_1, \ldots, v_n$ in
$G(T)$ such that for all $i\neq j$,
$\{v_i,v_{i+1}\} \in E'$ and $v_i \neq v_j$. The length of a path
$v_1,\ldots,v_n$ is $n-1$.

Formally, a  DTPRS $(\Rr,\Delta)$ is a {\em positive DTPRS} with set
of initial trees $\Init$, if:
\begin{itemize}
\item {\bf non-recursive-DTD:} the DTD in the static invariant $\Delta$ is
non-recursive. In particular, trees satisfying the DTD have depth
bounded by some $B>0$.
\item {\bf positive:} all guards in $\Rr$ and the data invariant in
  $\Delta$ are positive Boolean combinations of DTPs. The DTD in
  $\Delta$ is positive as well.
\item {\bf simple-path bounded:} there exists $K>0$ such that the
  length  of any simple path in any $T \in
  \Tt^\ast_{\Rr}(T_0)$ for any $T_0 \in
  \Init$, is bounded by $K$.
\end{itemize}

Notice that the third condition above implies
that all data trees have depth bounded by $K$. So we always
assume that $B \le K$. Notice also that in positive DTPRS, the data
value inequality is \emph{allowed} in DTPs, that is, we can state
that two data values are different.

The \emph{Play.com} example in Section~\ref{s:mo-example} satisfies the
first 2 conditions above.
However, in general, the third condition is not satisfied.  {\pid}s
can create create a long path: a cart can be linked to a product,
linked to another cart linked to another products etc.  So the number
of carts or the number of products needs to be bounded (unless a cart
can contain at most one product).  On the other hand, {\name} and
{Total} are fresh data values, they cannot be used as links. At last,
\cid\ can be used in different carts and orders, but as a cart or
order is associated to a unique customer, it cannot create long links.
More formally, if the system can handle only $C$ active carts at a
time (but the number of orders is unlimited), then the system has
simple paths bounded by $12C+7$. If there are at most $D$ different
products in the catalog, then the system has simple paths bounded by
$12D+7$. Finally, if each customer can have only one active cart at a
time (but she can have many orders), and each cart has at most one
product, then the system has simple paths bounded by $14$. Any of
these restrictions can be described using only positive rules. The
rest of the section is devoted to the proof of the following result:

\begin{theorem}\label{t.positive}
Given a positive DTPRS $(\Rr,\Delta)$, the pattern reachability and
the termination problem are decidable.
\end{theorem}

We prove Theorem~\ref{t.positive} by using the framework of
\emph{well-structured transition systems (WSTS)} \cite{acj96,FS01},
which has been applied to DTPRS \emph{without} data in
\cite{GMSZ08}. We recall briefly some definitions. A WSTS is a triple
$(S,\arr{},\preceq)$ such that $S$
is an (infinite) state space, $\preceq$ is a {\em
  well-quasi-ordering\footnote{A wqo $\preceq$ is a reflexive,
    transitive and well-founded relation with no infinite antichain.}
  (wqo for short)} on $S$, and $\arr{}$ is the transition relation on
$S$. It is required that  $\arr{}$ is \emph{compatible
  w.r.t.~$\preceq$}: for any $s,t,s'
\in S$ with $s \arr{} t$ and $s \preceq s'$, there exists
$t'\in S$ such that $s' \arr{} t'$ and $t \preceq
t'$.


Let $\Tt_{B,K}$ denote the set of data trees whose depths are
bounded by $B$ and lengths of simple paths are bounded by $K$. From
the definition of positive DTPRS, we know that
$\Tt^\ast_{\Rr}(\Init) \subseteq \Tt_{B,K}$.

In the following, we show Theorem~\ref{t.positive} by showing that
$(\Tt_{B,K}, \arr{}, \preceq)$ is a WSTS.

We define the binary relation $\preceq$ on $\Tt_{B,K}$.
Let $T_1=(V_1,E_1,\root_1,\ell_1),T_2=(V_2,E_2,\root_2,\ell_2) \in
\Tt_{B,K}$, then $T_1 \preceq T_2$ if there is an injective mapping
$\phi$ from $V_1$ to $V_2$ such that
\begin{itemize}
\item root preservation: $\phi(\root_1)=\root_2$,

\item parent-child relation preservation: $(v_1,v_2) \in E_1$ iff $(\phi(v_1),\phi(v_2)) \in
E_2$,

\item tag preservation: If $\ell_1(v) \in \Sigma$, then
$\ell_1(v)=\ell_2(\phi(v))$,

\item data value (in)equality preservation:
If $v_1,v_2 \in V_1$ and $\ell_1(v_1),\ell_1(v_2) \in
\Dd$, then $\ell_2(\phi(v_1)),\ell_2(\phi(v_2)) \in \Dd$, and
$\ell_1(v_1)=\ell_1(v_2)$ iff
$\ell_2(\phi(v_1))=\ell_2(\phi(v_2))$.
\end{itemize}

It is easy to see that $\preceq$ is reflexive and transitive, so it
is a quasi-order. In the following, we first assume that
$\preceq$ is a wqo on $\Tt_{B,K}$ and
show that $\arr{}$ is compatible with $\preceq$, in order to prove
Theorem~\ref{t.positive}. We
show in Section \ref{s:wqo} that $\preceq$ is indeed a wqo: for any infinite sequence
of data trees $T_0,T_1,\ldots \in \Tt_{B,K}$, there are $i<j$
such that $T_i \preceq T_j$.

\subsection{Well-structure of positive DTPRS}

Let $(\Rr,\Delta)$ be a positive DTPRS.

\begin{proposition}\label{p:compatibility}
  Let $T_1,T'_1,T_2 \in \Tt_{B,K}$, $T_1 \arr{R} T_2$ for some $R \in
  \Rr$, and $T_1 \preceq T'_1$.  Then there exists $T'_2 \in
  \Tt_{B,K}$ such that $T'_1 \arr{R} T'_2$ and $T_2 \preceq T'_2$.
\end{proposition}

\begin{proof}
Let $R=(L,G,\Qq,\Ff,\chi)$. Taking an injective mapping $\phi:T_1
\ra T'_1$ preserving the root, parent-child relation, tag, and
data (in)equality relation, and an injective matching $\psi: L \ra
T_1$ satisfying the data constraint $\cond$ of $L$, we have an
injective matching
$\phi \circ \psi : L \ra T^\prime_1$ which respects the
parent-child, tags and data (in)equality relation. Hence $\cond$
is satisfied by $\phi \circ \psi$ too. As $G$ is positive, if
$G$ is true at $T_1$ wrt. $\phi$, then it is true at $T^\prime_1$
wrt. $\phi \circ \psi$ as well. Applying the rule $R$ to
$T'_1$ wrt. $\phi \circ \psi$, we get a tree $T'_2$ such
that $T_2 \preceq T'_2$. As both the DTD and the data invariant in $\Delta$ are
positive and $T_2$ fulfills $\Delta$, so does $T'_2$. Thus
$T'_1 \arr{R} T'_2$.
\end{proof}

Consequently, we have shown that $\arr{}$ is compatible wrt.
$\preceq$ in $\Tt_{B,K}$, thus
$(\Tt_{B,K}, \arr{}, \preceq)$ is a WSTS.

In the following, we prove that $(\Tt_{B,K}, \arr{}, \preceq)$
satisfies some additional computability conditions, needed to show the
decidability of pattern reachability and termination.

First consider pattern reachability. To get the decidability of this
problem, from Theorem 3.6 in \cite{FS01}, we need to show that
$(\Tt_{B,K}, \arr{}, \preceq)$ has effective
pred-basis. A WSTS $(S,\arr{},\preceq)$ has \emph{effective
pred-basis} if there exists an algorithm that computes for any state
$s \in S$ the finite
 basis $pb(s)$ of the upward closed set $\uparrow \pred(\uparrow
 s)$. Here, $\uparrow I = \{s' \in S \mid \exists s \in I \text{
   s.t. } s \preceq s' \}$ denotes the upward closure of $I$ wrt. $\preceq$,
and $\pred(I)=\{s \in S \mid \exists t \in I, s \arr{} t\}$ the set of
immediate predecessors of states in $I$. A basis
of an upward-closed set $I$ is a minimal  set $I^b$ such that $I =
\bigcup_{x \in I^b} \uparrow x$. Recall that whenever $\preceq$ is a wqo, the
basis $I^b$ of an upward closed set $I$ is finite.

\begin{proposition}\label{p:eff-pre-basis}
$(\Tt_{B,K}, \arr{}, \preceq)$ has effective
pred-basis.
\end{proposition}

A solution for reachability of a given DTP $P$ from an initial set of
data trees $\text{Init}$ is obtained by backward exploration: we start
with $I^0$ as the set of data trees matching $P$ and satisfying
$\Delta$. Then compute iteratively the upward closed sets $I^{n+1} =
I^n \cup (\pred(I^n) \cap \Delta)$ by representing each set through its finite
basis. Since the sequence $I^n$ is increasing by construction, and
since $\preceq$ is a wqo, the sequence must be finite and termination
can be effectively tested. If $I^n=I^{n+1}$ then it suffices to check
whether $I^n \cap \Init =\emptyset$. Notice that we did not impose any
restriction on the set $\Init$ of the initial trees. We need to test
the existence of a data tree from $\Tt_{B,K}$ satisfying an (arbitrary) Boolean
combination on DTPs \emph{and} an (arbitrary) DTD. This problem is in
general undecidable \cite{Dav08}, but becomes decidable in the special
case where trees are of bounded depth \cite{Dav08}. Here we need to
talk in addition about trees from $\Tt_{B,K}$, but we can apply the
same proof ideas as in \cite{Dav08} in order to infer decidability.
%

Now consider the termination problem.
From Theorem 4.6 in \cite{FS01}, to show the decidability of
termination problem from a single initial tree $T_0$, it is
sufficient to show that $(\Tt^\ast_{\Rr}(T_0), \arr{R}, \preceq)$
has effective $\succ$, i.e. for each $T \in \Tt^\ast_{\Rr}(T_0)$,
the set $\succ(T):=\{T^\prime \mid T \arr{} T^\prime \}$ is
computable. Then we can compute the \emph{finite reachability tree}
starting with $T_0$: we compute trees $T$ s.t.~$T_0 \arr{*} T$ and
we stop whenever we find $T \preceq T'$ along some branch.

It is not hard to see that $\succ(T)$ contains only a finite number
of equivalence classes induced by the quasi-order $\preceq$. Since
the DTPRS $(\Rr, \Delta)$ is not able to distinguish between two
distinct data trees belonging to the same equivalence class, by
selecting one data tree from each equivalence class, we can get a
finite representation of $\succ(T)$, therefore,
$(\Tt^\ast_{\Rr}(T_0), \arr{}, \preceq)$ has effective $\succ$.

\subsection{Tree Decompositions}\label{s:wqo}

\def\Gg{\mathcal{G}}
\def\insubg{\sqsubseteq}

In order to prove that $\preceq$ is a wqo over $\Tt_{B,K}$, we first
represent a data tree $T$ as a (labeled) undirected graph
$G_{\ell}(T)$, then we encode $G_{\ell}(T)$ into a  tree
(without data) of \emph{bounded depth} using the concept of tree
decompositions. Define a binary relation $\le$ on (labeled) trees of
bounded depth as follows: $T_1 \le T_2$ if there is an injective
mapping from $T_1$ to $T_2$ preserving the root, the tags, and the
parent-child relation. It is known that $\le$ is a wqo on
labeled trees of bounded depth \emph{without data} \cite{GMSZ08}.

Let $\Gg_K$ be the set of labeled graphs with the lengths of all
simple paths bounded by $K$. We show that $\preceq$ on $\Tt_{B,K}$
corresponds to the induced subgraph relation (formally defined
later) on $\Gg_K$, and the fact that $\le$ is a wqo for labeled
trees of bounded depth implies that the induced subgraph relation is
a wqo on $\Gg_K$.

Given a data tree $T=(V,E,root,\ell) \in \Tt_{B,K}$, \emph{the
labeled undirected graph representation $G_{\ell}(T)$  of $T$} is
obtained from $G(T)$, the graph associated to $T$, by adding labels
encoding information of data tree nodes (tag, depth $\ldots$).
Formally, $G_{\ell}(T)$, is a $\left((\Sigma \cup \{\#\}) \times
[B+1]\right) \cup \{\$\}$-labeled (where $[B+1]=\{0,1,\cdots,B\}$)
undirected graph $(V^\prime, E^\prime, \ell^\prime)$ defined as
follows,
\begin{itemize}
\item $V^\prime = V \cup \{\ell(v) \mid v \in V, \ell(v) \in \Dd\}$,

\item $E^\prime=E \cup \{\{v,d\} \mid v \in V, \ell(v)=d \in \Dd\}$,

\item If $\ell(v) \in \Sigma$, then $\ell^\prime(v)=(\ell(v),i)$,
otherwise, $\ell^\prime(v)=(\#,i)$, where $i$ is the depth of $v$ in
$T$. In addition, $\ell^\prime(d)=\$$ for each $d \in V^\prime \cap
\Dd$.
\end{itemize}

Let $\Sigma_G$ denote $\left((\Sigma \cup \{\#\}) \times
[B+1]\right) \cup \{\$\}$.

For $\Sigma_G$-labeled graphs, we define the induced subgraph relation
as follows. Let $G_1=(V_1,E_1,\ell_1),G_2=(V_2,E_2,\ell_2)$ be two
$\Sigma_G$-labeled graphs, then $G_1$ is an \emph{induced subgraph}
of $G_2$ (denoted $G_1 \insubg G_2$) iff there is an injective
mapping $\phi$ from $V_1$ to $V_2$ such that
\begin{itemize}
\item label preservation: $\ell_1(v_1)=\ell_2(\phi(v_1))$ for any $v_1 \in
V_1$,

\item edge preservation: let $v_1,v^\prime_1 \in V_1$, then $\{v_1,v^\prime_1\} \in
E_1$ iff $\{\phi(v_1),\phi(v^\prime_1)\} \in E_2$.
\end{itemize}

From the definition of the labeled graph representation of data
trees, it is not hard to show that the induced subgraph relation
$\insubg$ corresponds to the relation $\preceq$ on data trees.

\begin{proposition}\label{prop-data2graph}
Let $T_1,T_2 \in \mathcal{T}_{B,K}$, then $T_1 \preceq T_2$ iff
$G_{\ell}(T_1) \insubg G_{\ell}(T_2)$.
\end{proposition}


Now we show how to encode any $\Sigma_G$-labeled graph belonging to
$\Gg_K$ into a labeled tree of bounded depth by using tree
decompositions.

Let $G=(V,E,\ell)$ be a connected $\Sigma_G$-labeled graph, then a
\emph{tree decomposition} of $G$ is a quadruple $T=(U,F,r,\theta)$
such that:
\begin{itemize}
\item $(U,F,r)$ is a tree with the tree domain $U$, the parent-child relation $F$, and the root $r \in U$,

\item $\theta: U \rightarrow 2^V$ is a labeling function attaching each
node $u \in U$ a set of vertices of $G$,

\item For each edge $\{v,w\} \in E$, there is a node $u \in U$ such
that $\{v,w\} \subseteq \theta(u)$,

\item For each vertex $v \in V$, the set of nodes $u \in U$ such
that $v \in \theta(u)$ constitutes a connected subgraph of $T$.
\end{itemize}

The sets $\theta(v)$ are called the \emph{bags} of the tree
decomposition.

The \emph{depth} of a tree decomposition $T=(U,F,r,\theta)$ is the
depth of the tree $(U,F,r)$ and the \emph{width} of $T$ is defined
as $\max\{|\theta(u)|-1 \mid u \in U\}$. The \emph{tree-width} of a
graph $G=(V,E)$ is the minimum width of tree decompositions of $G$.
For a tree decomposition of width $K$ of a graph $G$, without loss
of generality, we assume that each bag is given by a sequence of
vertices of length $K+1$, $v_0\dots v_K$, with possible repetitions,
i.e. possibly $v_i = v_j$ for some $i,j: i \ne j$ (tree
decompositions in this form are sometimes called ordered tree decompositions).


\begin{theorem}[\cite{NM06,BC09}]\label{t:path-2-tree-decom}
If $G \in \Gg_K$, then $G$ has a tree decomposition with both depth
and width bounded by $K$.
\end{theorem}

\begin{proof}
Let $G=(V,E,\ell) \in \Gg_K$ and $T=(V,E_T,r)$ be a
depth-first-search tree of $G$ with $r \in V$ as the root. Then $T$
is of depth at most $K$. For each $v \in V$, let $\theta(v)$ be the
union of $\{v\}$ and the set of all ancestors of $v$ in $T$, then
$(V,E_T,r,\theta)$ is a tree decomposition of $G$ of depth at most
$K$ and width at most $K$.
\end{proof}

As a matter of fact, the converse of
Theorem~\ref{t:path-2-tree-decom} holds as well.

\begin{proposition}
If $G$ has a tree decomposition of width $\le A$ and depth $\le B$,
then the length of any simple path of $G$ is bounded by
$(A+2)^{B}+\sum_{1 \le i \le B}(A+2)^i$.
\end{proposition}

So generally speaking, for a class of graphs,  all simple paths are
length-bounded for each graph in the class iff there is a tree
decomposition of bounded depth and width for each graph in the
class.

%
%
%
%
%
%


Now we describe how to encode labeled graphs by labeled trees using
tree decompositions.

Let $G=(V,E,\ell) \in \Gg_K$ be a $\Sigma_G$-labeled graph, and
$T=(U,F,r,\theta)$ be a tree decomposition of $G$ with width $K$ and
depth at most $K$. Remember that each $\theta(u)$ is represented as
a sequence of exactly $K+1$ vertices, and $[K+1]=\{0,\dots,K\}$.
Define
\[\Sigma_{G,K}:=(\Sigma_G)^{K+1} \times 2^{[K+1]^2} \times
2^{[K+1]^2} \times 2^{[K+1]^2}.\]
We transform $T=(U,F,r,\theta)$ into a $\Sigma_{G,K}$-labeled tree
$T^\prime=(U,F,r,\eta)$, which encodes in a uniform way the
information about $G$ (including edge relations and vertex labels).
$\eta: U \rightarrow \Sigma_{G,K}$ is defined as follows. Let
$\theta(u)=v_0 \dots v_K$, then $\eta(u)=(\ell(v_0)\dots \ell(v_K),
\bar{\lambda})$, where
$\bar{\lambda}=(\lambda_1,\lambda_2,\lambda_3)$,
\begin{itemize}
\item $\lambda_1=\{(i,j) \mid 0 \le i, j \le K, v_i=v_j\}$,
\item $\lambda_2=\{(i,j) \mid  0 \le i, j \le K, \{v_i,v_j\} \in E\}$,
\item If $u=r$, then $\lambda_3=\emptyset$,
otherwise let $u^\prime$ be the parent of $u$ in $T$ and
$\theta(u^\prime)=v^\prime_0 \cdots v^\prime_K$, then $\lambda_3 =
\{(i,j) \mid 0 \le i, j \le K, v^\prime_i=v_j\}$.
\end{itemize}
\vspace*{-3mm}



\subsection{Well-quasi-ordering for data trees}\label{s:wqo}

The encoding of labeled graphs into labeled trees establishes a
connection between the wqo $\le$ of labeled trees and the induced
subgraph relation ($\insubg$) of labeled graphs.

\begin{proposition}\label{prop-graph2tree}
Let $G_1,G_2$ be two $\Sigma_G$-labeled graphs with tree-width
bounded by $K$, and $T_1,T_2$ be two tree decompositions of width
$K$ of resp. $G_1,G_2$, then the two $\Sigma_{G,K}$-labeled trees
$T^\prime_1,T^\prime_2$ obtained from $T_1,T_2$ satisfy that: If
$T^\prime_1 \le T^\prime_2$, then $G_1 \insubg G_2$.

\end{proposition}

\begin{proof}
Let $G_i=(V_i,E_i,\ell_i)$, $T_i=(U_i,F_i,r_i,\theta_i)$ and
$T^\prime_i=(U_i,F_i,r_i,\eta_i)$($i=1,2$). Suppose that $T^\prime_1
\le T^\prime_2$. Then there is an injective mapping $\phi$ from
$U_1$ to $U_2$ preserving the root, the parent-child relation and
the node-labels.

Define an injective mapping $\pi: V_1 \rightarrow V_2$ as follows:
\begin{quote}
For $v \in V_1$, select some $u \in U_1$ such that
$\theta_1(u)=v_0\dots v_K$ and $v=v_i$ for some $i$. Writing
$\theta_2(\phi(u))=v^\prime_0\dots v^\prime_K$, we let
$\pi(v)=v^\prime_i$.
\end{quote}

First we show that $\pi$ does not depend upon the choice of $i$ such
that $v=v_i$, neither on the choice of $u \in U_1$ such that $v \in
\theta_1(u)$. The former holds because $\eta_1(u)=\eta_2(\phi(u))$
(and in particular the component $\lambda_1$ is preserved), hence if
$v_i=v_j$, then we also have $v'_i=v'_j$.

For the latter, notice that the $\lambda_3$ component of
$\eta_1(u)=\eta_2(\phi(u))$ is preserved, hence the choice of $u$ or
of its father is irrelevant. Now, the set $\{u \in U_1 \mid v \in
\theta_1(u)\}$ is a connected subgraph of $T_1$ by definition of
tree decomposition, hence $\pi$ does not depend upon the choice of
$u \in U_1$.

Now we show that $\pi$ is injective. Let $v_2$ be a vertex of $G_2$.
Because of the preservation of $\lambda_1$, no two different
vertices $v,v'$ of $G_1$ with  $v,v' \in \theta(u)$ can satisfy
$\pi(v)=\pi(v')=v_2$. Because of the preservation of $\lambda_3$, no
two different vertices $v,v'$ with $v\in \theta(u)$ and $v' \in
\theta(u')$ with $u$ father of $u'$ can satisfy
$\pi(v)=\pi(v')=v_2$. Again, as the set $\{u \in U_1 \mid v_2 \in
\theta(u)\}$ is a connected subgraph of $T_2$, it means that $\pi$
is injective.

We finish the proof by showing that $\pi$ preserves the node-labels
and edge relations.

\textbf{Node-label preservation}: Suppose $\pi(v)=v^\prime$. Then
there exists some $u \in U_1$ such that $\theta_1(u) = v_0 \cdots
v_k$, $v=v_i$ for some $i$, $\theta_2(\phi(u)) = v^\prime_0 \cdots
v^\prime_k$, and $v^\prime=v^\prime_i$. Since
$\eta_1(u)=\eta_2(\phi(u))$, $\ell_1(v_0) \dots
\ell_1(v_k)=\ell_2(v^\prime_0)\dots \ell_2(v^\prime_k)$, it follows
that $\ell_1(v)=\ell_1(v_i)=\ell_2(v^\prime_i)=\ell_2(v^\prime)$.

\smallskip

\textbf{Edge relation preservation}: We show that $\{v,w\} \in E_1$
iff $\{\pi(v),\pi(w)\} \in E_2$ for any $v,w \in V_1$.

If $\{v,w\} \in E_1$, there exists $u \in U_1$ such that $\theta_1(u)=v_0 \cdots v_k$, $v=v_i$ and $w=v_j$ for some
$i,j$. So $(i,j) \in \lambda_2(u)$ in $T^\prime_1$. Then $(i,j) \in
\lambda_2(\phi(u))$. Let $\theta_2(\phi(u))=v^\prime_0 \cdots
v^\prime_k$, then $\{v^\prime_i,v^\prime_j\} \in E_2$. Consequently
$\{\pi(v),\pi(w)\}=\{v^\prime_i,v^\prime_j\} \in E_2$.

If $\{\pi(v),\pi(w)\} \in E_2$, then there exists $u^\prime \in U_2$
such that $\pi(v),\pi(w) \in \theta_2(u^\prime)$. Without loss of
generality, we can choose $u^\prime$ at minimal depth such that
$\pi(v),\pi(w) \in \theta_2(u^\prime)$. It means that for instance,
the father $u^{\prime\prime}$ of $u^\prime$ satisfies $\pi(v) \notin
\theta_2(u^{\prime\prime})$. Since
$U^\prime_2=\{u^{\prime\prime\prime} \in U_2 \mid \pi(v) \in
\theta_2(u^{\prime\prime\prime})\}$ is connected, it means that
$U^\prime_2$ is entirely contained in the subtree rooted at
$u^\prime$. By contradiction, if there does not exist $u \in U_1$
such that $\phi(u)=u'$, then $\phi(U_1) \cap U^\prime_2 =
\emptyset$. On the other hand, according to the definition of $\pi$,
there is $u \in U_1$ such that $v \in \theta_1(u)$ and $\pi(v) \in
\theta_2(\phi(u))$. So $\phi(u) \in \phi(U_1) \cap U^\prime_2$, a
contradiction. Thus there is $u \in U_1$ such that
$u^\prime=\phi(u)$. Let $\theta_1(u) = v_0 \dots v_k$ and
$\theta_2(u^\prime) = v^\prime_0 \dots v^\prime_k$, by injectivity
of $\pi$, we have $\pi(v)=v^\prime_i$, $\pi(w)=v^\prime_j$, $v=v_i$,
$w=w_j$ for some $i,j$. Then $(i,j) \in \lambda_2(u^\prime) =
\lambda_2(u)$, which proves that $\{v,w\}$ is an edge of $G_1$.
\end{proof}

Now we are ready to show that $\preceq$ is a wqo for $\Tt_{B,K}$.

Let $T_0,T_1,\dots$ be an infinite sequence of data trees from
$\Tt_{B,K}$. Consider the infinite sequence of
$\Sigma_{G,K}$-labeled trees $T^\prime_0,T^\prime_1,\dots$ obtained
from the tree decompositions (with width $K$ and depth at most $K$)
of graphs $G_{\ell}(T_0),G_{\ell}(T_1),\dots$. Then there are $i,j:
i < j$ such that $T^\prime_i \le T^\prime_j$, because $\le$ is a wqo
for labeled trees of depth at most $K$. So $G(T_i) \insubg G(T_j)$
from Proposition~\ref{prop-graph2tree}, and $T_i \preceq T_j$ from
Proposition~\ref{prop-data2graph}. We thus prove following theorem.

\begin{theorem}
$\preceq$ is a well-quasi-ordering over $\Tt_{B,K}$.
\end{theorem}

\section{Verification of temporal properties}

Until now we considered only two properties for static analysis:
termination and pattern reachability. (Non-)reach\-abi\-lity of a DTP can be
expressed easily in Tree-LTL \cite{ASV08}, which corresponds roughly
to linear time temporal logics where atomic propositions are
DTPs\footnote{Such formulas use actually free variables in patterns,
  which are then quantified universally. This is consistent with the
  approach of testing whether a model satisfies the negation of a
  formula.}. We show in this section that allowing for runs of
unbounded length makes the validation of (even simple) Tree-LTL
properties undecidable, even without data:

\begin{proposition}\label{p:ltl}
  It is undecidable whether a TPRS with initial (data-free) tree $T_0$ satisfies
  a given Tree-LTL formula.
\end{proposition}

The fact that Tree-LTL is undecidable does not disallow us to verify
quite complicated properties. We show on the \emph{Play.com} example
how to proceed: it suffices to encode in the system the property we
want to check with additional tags and check for pattern
reachability. For example, suppose that we want to verify whether a
customer can add some product after the bill was processed. For
that, we add new tags $\#$, $1,2,3$ to the alphabet, and we add one
child to \play\ labeled by $1$ in the initial tree. We add one rule
which checks that the additional tag is 1 and selects one cart in
the payment mode. The outcome of the rule is to change the tag from
$1$ to $2$, and to append $\#$ as child of the selected cart.  We
add another rule which checks that the tag is 2. The outcome of the
rule is to change the tag from $2$ to $3$, and to append a new
product with one item in the inventory below \pcat\, with a special
marker $\#$ as brother of \pid . Now one can reach in the new system
a tree with a cart marked $\#$ and with a product with \pid\ $X$
such that there exist a Product with \pid\ $X$ in the PCatalog which
is marked by $\#$ iff a customer can add some product after the bill
was generated in the original system. The former property is a
pattern reachability problem, which we proved to be decidable.

\section{Bounded model-checking DTPRS and recursion-free GAXML}

\def\init{\text{\emph{Init}}}
\def\nexptime{\textsf{NexpTime}}

Recall that \cite{ASV08} shows that the largest decidable fragment of
GAXML that can be model-checked w.r.t.~Tree-LTL properties is the
recursion-free one. Absence of recursion in GAXML roughly means (1) disallowing
recursive DTDs (as we do here) and (2) imposing that no function is
called more than once, on any execution path. On the other hand, one
can use negated DTPs in this fragment.

In this section we consider \emph{bounded model checking} for DTPRS:
Given a DTPRS $(\Rr,\Delta)$, a set of initial trees $\init$, a DTP
$P$ and a bound $N$ (encoded in unary) we ask whether there is some
$T_0$ satisfying $\init$ and some $T$ s.t.~$P$ matches $T$ and $T_0
\arr{\le N} T$. We show the following result:

\begin{theorem}\label{t:bmc}
  Bounded model-checking for DTPRS is \\
  $\nexptime$-complete.
\end{theorem}

Theorem~\ref{t:bmc} can be actually extended to bounded
model-checking Tree-LTL properties. Bounded model-checking of a
Tree-LTL formula $\varphi$ with a bound $N$ is the problem checking
whether a counter-example for $\varphi$ holds  in $\le N$-steps.
For instance, bounded model-checking for $G \neg P$ with a bound $N$
is to check whether the DTP $P$ can be reached in $\le N$ steps.


For the upper bound we show how to encode a DTPRS $(\Rr,\Delta)$
with the given bound $N$ into a recursion-free GAXML system, and use
the upper bound provided by \cite{ASV08}. We recall that
\cite{ASV08} provides a \emph{simply} exponential bound in the
number of transition steps of the recursion-free GAXML system. Also
notice that the DTPRS in  Theorem \ref{t:bmc} are not supposed to be
positive -- the lower bound  relies on negations of DTPs.




The basic idea of the reduction from bounded model-checking DTPRS to
recursion-free GAXML is the following: we ``guess''  on-the fly the
rules $R_1,R_2 ,\ldots,R_M$, $M \le N$, that are applied on a
successful path of the DTPRS, and use function labels for
pinpointing the nodes used by the matching of the corresponding
locators. Suppose that nodes of locators have identifiers. A node
having a child (leaf) labeled by the function call $! (i,R_i,w)$
with $i \le M$, $R_i \in \Rr$ and $w$ an identifier within the
locator $L_i$ of rule $R_i$, is ``guessed'' to correspond to node
$w$ in the matching of the locator $L_i$ when applying rule $R_i$.
Notice that node can have several function calls $(i,R_i,v),
(j,R_j,w)$ attached to it (but then, $i \not= j$, since the matching
of $L_i$ should be injective).

We use the DTD in the invariant $\Delta$ in order to ensure that (1)
each of the (polynomially many) function labels $!  (i,R_i,w)$
occurs at most once at any time point, and (2) we use GAXML
call/return guards for ensuring that calls related to rule $R_i$ are
only performed after all function calls $(j,R_j,v)$ with $j<i$, have
been completed, for each $i$. Checking (2) is done by forbidding the
presence of function calls $!(j,R_j,v)$ and $?(j,R_j,v)$ with $j<i$,
whenever $(i,R_i,w)$ is called. Similarly, when the result of a call
$? (i,R_i,w)$ is returned, we forbid that it contains some label
$!(j,R_j,v)$ or $?(j,R_j,v)$ with $j<i$.

Applying rule $R_i$ means calling all functions $! (i,R_i,w)$
one-by-one (say in DFS order) and performing the associated actions.
When we call the first function with index $i$, its guard also
checks that the locator $L_i$ was properly guessed. A rename action
must be simulated, since GAXML has no renaming facility: a call
$!(i,R_i,w)$ with $w$ labeled by $\ren_b$ in $L_i$ has as effect to
attach label $b_i$ as a child (leaf) of node $w$. Checking which is
the current label of a node is done by using negative guards: we
look for a child $b_i$ that has no sibling $c_k$ with $k>i$. A
delete action must be simulated, too, since GAXML has no deletion.
A call $!(i,R_i,w)$ with $w$ labeled by $\del$ in $L_i$ has as
effect to return a node with tag \emph{del}. This might be
syntactically inconsistent if the current node is a data
node. However, standard encoding tricks can remedy this problem. For
simplicity, let us assume that the tag
\emph{del} is always appended as a sibling of the node that is supposed
to be deleted in DTPRS.

Finally, the append action is done as in GAXML, by performing the
query and attaching the result. Here, we need to take care about the
nodes that added via a forest $F \in \Ff$, resp.~a query. For such
nodes, we must ``guess'' some attached function calls. In both cases
we use external GAXML functions: for example, we can simulate the
addition of an annotated copy of $F$ via an external call, and use
$\Delta$ for checking that the right $F$ was added. For query answers
we may split a query $Q$ in polynomially many copies, where for some
of the copies, the $\head$ has attached external function calls. Their
role is to generate (sets) of functions of type $!(i,R,v)$ attached to
nodes in $\head$.


The last point is that we need to adapt the GAXML upper bound in
order to take care of nodes marked by \emph{del}: this can be done
e.g.~by extending the notion of matching tree patterns in such a way
that none of the nodes to which tree patterns are matched, nor their
ancestors, are allowed to have a child labeled  by the tag
\emph{del}. It can be easily checked that the complexity checking a
Tree-LTL formula for recursion-free GAXML still holds with this
extended notion of matching. The reason is that the proof is based
on small models obtained by taking tree prefixes of the bigger
model. Obviously, the absence of children with tag \emph{del} is
preserved by taking tree prefixes.

The lower bound is adapted from the 2-$\nexptime$ lower bound proof for
recursion-free GAXML. We only recall the rough idea here.

The main ingredient of the proof is to create/check lists of length
$2^n$. This is done using data values, similarly as in the proof of
Theorem \ref{t:fresh}. A ``list'' of length $k$ corresponds to a tree
of depth 2, where each node at depth one has 2 children, with distinct
data values $d_i,d_{i+1}$. If each data value $d_i$ occurs twice (except for
$d_1$ and $d_{k+1}$, which occur only once) we get a linear order,
i.e.~a list. Using $n$ queries we can compute $n$ steps of transitive
closure and thus verify that $k=2^n$. Obviously, this suffices for
encoding a $(2^n \times 2^n)$ tableau representing a computation of a
$2^n$-time bounded TM. Details are fairly easy to complete.

\section{Conclusion}

In this paper, we defined a rich class of systems describing active
documents, possibly with recursive calls. We show that this class of
systems is easy to use and powerful, demonstrating it on the
MailOrder example.  We studied the boundary of decidability for
different properties and restrictions of the active documents.
Namely, we show that termination from one document and pattern
reachability from a set of documents are both decidable for {\em
positive} DTPRSs, which are DTPRSs where the DTD is non-recursive,
there is no negative guard or data invariants, and {\em
  simple paths} are bounded. We showed that without these restrictions, the problem is undecidable. We also show that the
problem is undecidable for more complex properties (Tree LTL or
termination from a set of active documents). Nevertheless, we also
demonstrate on the MailOrder example that one can find bugs with our
method.

Compared with GAXML \cite{ASV08}, the respective restrictions used
to get decidability (positiveness and non-recursion) are
incomparable. We showed however a reduction from (not necessarily
positive) rewriting-length bounded DTPRS to recursion-free GAXML.

Considering further work it seems possible to get decidability
results for another (incomparable) class of systems, namely DTPRSs
whose new data variables do not get mutually distinct fresh values,
but possibly arbitrary data values.


\bibliographystyle{abbrv}
\bibliography{DTPRS}

\newpage

\

\newpage

\appendix

\emph{Proof of Proposition 2:}

  In both cases we encode a 2-counter machine with counters $a,b$. In
  the first one, a
  configuration $(q,n_a,n_b) \in Q \times \Nat \times \Nat$ is encoded
  by a tree with root labeled
  $q$ and two subtrees, one of the form $a^{n_a}a_\$$, and the other
  of the form $b^{n_b}b_\$$. E.g.~a zero test on the first counter
  corresponds to checking that the root has a child labeled
  $a_\$$. Decrementing the first counter in state $q$ (and going to
  state $q'$) is done using the locator $[q](-[a]([a_\$]))$, where the
  additional labels are: $\ren_{q'}$ for the root, $\ren_{a_\$}$ for
  the $a$-node and $\del$ for the $a_\$$-node.

With non-recursive DTDs we can encode a configuration  $(q,n_a,n_b)$ by a
tree of depth one, with root labeled by $q$, and $n_a$ ($n_b$, resp.)
leaves labeled by $a$ ($b$, resp.). The zero test is now done using a
negative guard (e.g.~"no $a$-leaf") or a negative invariant. In the
latter case we split a transition in 2 steps: first we relabel the
root by a transition from that state; second, we perform the
corresponding rewriting as before. The invariant states that whenever
the root is labeled by a transition corresponding to a zero test of
counter $c$, the
tree has no $c$-leaf ($c \in \{a,b\}$).
\qed

\medskip

\medskip

\emph{Proof of Proposition 7:}

It is sufficient to consider $\min(\pred(\uparrow T))$, the set of
minimal elements wrt. $\preceq$ in $\pred(\uparrow T)$, for each tree
$T \in \Tt_{B,K}$.

Fix a rule $R=(L,G,\Qq,\Ff,\chi)$ with
\[L=(V_L,E_L,root_L,\ell_L,\tau_L,cond_L,\ell^\prime_L)\]
such that $\ell^\prime_L$ attaches additional labels
$\{\app,\ren_a,\del\}$, $\Qq=\{Q_1,\dots, Q_m\}$ ($Q_i=\body_i \rsa
\head_i$), and $\Ff=\{F_1,\dots,F_n\}$.

Let $T_1=(V_1,E_1,root_1,\ell_1) \in \min(\pred(\uparrow T))$, then
\begin{quote}
$\exists T^\prime$ such that $T_1 \arr{R} T^\prime$ wrt. some
$\phi:L \ra T_1$ and $T \preceq T^\prime$ via some $\psi: T \ra
T^\prime$. \hfill ($\ast$)
\end{quote}

In the following, we show that the size of $T_1$ (number of nodes)
is bounded by the following constant (we actually show even more, by
exhibiting $T_1$ satisfying $\Delta$):
\[B^2\left((|\Sigma|+1)\max(\Delta)\right)^B\left(|L|+|G|+|T|\max
\limits_{i}|Q_i|\right).\]

Thus a finite basis, which is a finite subset of $\min(\pred(\uparrow
T))$, is computable.

Let $T^\prime=(V^\prime,E^\prime,root^\prime,\ell^\prime)$. Then
$V^\prime$ consists of four disjoint subsets,
\begin{itemize}
\item $V^\prime_1 = \{\phi(v) \mid v \in V_L, v \mbox{ not labeled by \del} \}$,
\item $V^\prime_2 = node^{-1}(V^\prime_1)$, where
$node^{-1}(V^\prime_1)$ is the set of nodes $w \in V_1 \setminus
\phi(V_L)$ such that the lowest ancestor of $w$ in $\phi(V_L)$ is in
$V^\prime_1$.
\item $V^\prime_3$ contains distinct copies of $F_j$, excluding the leaves labeled by those $Q_i$,
\item $V^\prime_4$ contains distinct copies of the nodes of the forest $Q_i(T_1)$, one for each node labeled by $Q_i$ in each copy of $F_j$.
\end{itemize}

The node set of $T_1$ consists of $V^\prime_1$, $V^\prime_2$,
$V_3=\{\phi(v) \mid v \in V_L, v \mbox{ labeled by \del}\}$, and
$V_4=node^{-1}(V_3)$.

Now we consider an upper bound on the size of $T_1$ that are
sufficient to allow $T_1$ satisfying ($\ast$),
\begin{itemize}
\item To guarantee the matching $\phi$ from $L$ to $T_1$: \\
The nodes in $V^\prime_1 \cup V_3=\phi(V_L)$ and all their ancestors
in $T_1$ are
sufficient.\\
Note that in $L$, ancestor relations $||$ may occur, so the
inclusion of the ancestors of nodes in $V^\prime_1 \cup
V_3=\phi(V_L)$ is necessary.\\
Size: $B|\phi(V_L)|=B|L|$;

\item To witness that $G$ is satisfied over $T_1$ wrt. $\phi$: \\
$G$ is a positive Boolean combination of DTPs. To witness the
satisfaction of each DTP $P_i$ in $G$, we need keep a matching
$\phi_i$ from $P_i$ to $T_1$ and all the ancestors of nodes of
$\phi_i(P_i)$ in
$T_1$. \\
Size: $B|G|$;

\item To guarantee that $T \preceq T^\prime$:
\begin{itemize}
\item Keep $(V^\prime_1 \cup V^\prime_2) \cap \psi(V_T)$ and all their ancestors in $T_1$,
\item At most $|T|$ instantiations of $\head_i$ on $T_1$ by
matchings from $\body_i$ to $T_1$ wrt.~$\phi$ are sufficient. The
ancestors of all the nodes of $T_1$ in these instantiations should
be preserved as well.
\end{itemize}
Size: $B|T|+B|T||body_i| \le B|T|\max \limits_{i}|Q_i|$.

\item Finally, to satisfy the DTD in $\Delta$, $T_1$ should be completed
into a data tree of size at most (c.f.~\cite{ASV08}) \[B \cdot
\left(|(\Sigma|+1) \max(\Delta)\right)^B |T_1|,\] where
$\max(\Delta)$ is the maximum integer used in the definition of DTD
in $\Delta$.
\end{itemize}

Thus a sufficient upper bound for the size of $T_1$ is
\[B^2\left((|\Sigma|+1)\max(\Delta)\right)^B\left(|L|+|G|+|T|\max
\limits_{i}|Q_i|\right).\]
\qed

\medskip

\emph{Proof of Proposition 10.}

Let $G=(V,E)$ and $T=(W,F,r,\theta)$ be a tree decomposition of $G$
of width at most $A$ and depth at most $B$.

Let $P=v_1 \cdots v_n$ be a path in $G$, and $w_1 \cdots w_n$ be a
trace of $P$ in $T$ such that $v_i \in \theta(w_i)$, $w_i=w_{i+1}$
or there is a path in $T$ from $w_i$ to $w_{i+1}$ such that for each
$w \ne w_{i+1}$ on the path, $v_i \in \theta(w)$.
%
%

Because all bags are of size at most $A+1$, each bag can only occur
at most $A+1$ times on the sequence $w_1\cdots w_n$.

Let $B_0$ be the minimal depth of $w_i$'s. Then there is only one
bag at depth $B_0$, say $w$, occurring on the sequence $w_1 \cdots
w_n$.

Let $w_{i_1},\cdots,w_{i_l}$ ($l \le A+1$, $i_j < i_{j+1}$) be all
the occurrences of $w$ on the sequence $w_1 \cdots w_n$. Then all
the bags on each sub sequence $w_{i_j+1}w_{i_j+2} \cdots
w_{i_{j+1}-1}$ is at depth no less than $B_0+1$. By induction
hypothesis, each subsequence $w_{i_j+1} w_{i_j+2}\cdots
w_{i_{j+1}-1}$ is of length at most
\[(A+2)^{B-B_0-1}+\sum \limits_{1 \le i \le B-B_0-1} (A+2)^i,\]

thus
\[\begin{array}{ccl}
    n & \le & l + (l+1) \left((A+2)^{B-B_0-1}+\sum \limits_{1
\le i \le B-B_0-1} (A+2)^i\right)  \\
    & \le & (A+2) \left(1+(A+2)^{B-1}+\sum
\limits_{1 \le i \le B-1} (A+2)^i\right)\\
    & = & (A+2)^B + \sum \limits_{1 \le i \le B} (A+2)^{i}.
  \end{array}
\]
\qed

\newpage

\emph{Proof of Proposition 13:}

  We give a reduction from the \emph{exact} reachability problem, i.e. checking whether a tree
  $T_2$ can be reached \emph{exactly} from a tree $T_1$ via a TPRS
  $\Rr$. This problem was shown to
  be undecidable in \cite{GMSZ08} by a reduction from the reachability
  problem on reset Petri nets\footnote{Although the TPRS model used in
  \cite{GMSZ08} is slightly more general than the present model, it is
easily seen that in the reduction in \cite{GMSZ08} we use the TPRS
model presented here.}.

We reduce the exact reachability problem for $\Rr$ to checking the
Tree-LTL formula $\f={\bf  G}(P_1 \rightarrow{\bf  F} \, P_2)$
($P_1$ and $P_2$ are DTPs) for a TPRS
$\Rr'$ and initial tree $T_0$.

Let $\S$ be the set of tags of $\Rr$, and let $\S_m$ be a disjoint
copy of $\S$. We will use five new tags $root,\alpha, \beta, \gamma,
\delta \notin \Tt$, thus $\Tt'=\Tt \cup \S_m \cup
\{root,\alpha,\beta,\gamma,\delta\}$.

The starting tree $T_0$ consists of a root with one child labeled
$\alpha$, and one other child tree that equals $T_1$.

The TPRS $\Rr'$ consists of $\Rr$ (adapted in order to take the
additional root into account), plus two kinds of new rules (with
empty guards):\\[.1cm]

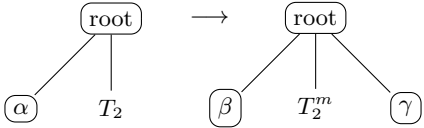
\begin{minipage}[h]{.2\linewidth}
 \begin{tikzpicture}[scale=.8,level 2/.style={sibling distance = 1cm}]
\node [rounded corners, rectangle, draw] {root} child {node[rounded
corners, rectangle, draw] {$\alpha$}
       }
child  {node {$T_2$}
       }
child[grow = right,white,level distance=12ex] {node[black]
{$\longrightarrow$}};
\end{tikzpicture}
 \end{minipage} \hspace{6ex}
 \begin{minipage}[h]{.2\linewidth}
    \begin{tikzpicture}[scale=.8,level 2/.style={sibling distance = 1cm}]
\node [rounded corners, rectangle, draw] {root} child {node[rounded
corners, rectangle, draw] {$\beta$}
       }
child  {node {$T^m_2$}
       }
child {node [rounded corners, rectangle, draw]{$\gamma$}
       }
;
\end{tikzpicture}
 \end{minipage}

 In the rule above, $T^m_2$ means that we use the tag copy $\S_m$ for
 $T_2$. Technically, the rule renames $\alpha$ by $\beta$, renames each tag $a \in \S$ in $T_2$ by $a_m \in
 \S_m$, and appends a new node with tag $\gamma$.

The second type of additional rules is the following (here, $a \in
\Sigma$ parametrizes the rule):

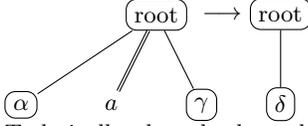
\begin{minipage}[h]{.2\linewidth}
 \begin{tikzpicture}[scale=.8,level 2/.style={sibling distance = 1cm}]
\node [rounded corners, rectangle, draw] {root} child {node[rounded
corners, rectangle, draw] {$\alpha$}
       }
child  {node {$a$} edge from parent[double]
       }
child {node [rounded corners, rectangle, draw]{$\gamma$}
      }
child[grow = right,white,level distance=8ex] {node[black]
{$\longrightarrow$}};
\end{tikzpicture}
 \end{minipage} \hspace{10ex}
 \begin{minipage}[h]{.2\linewidth}
    \begin{tikzpicture}[scale=.8,level 2/.style={sibling distance = 1cm}]
\node [rounded corners, rectangle, draw] {root} child  {node
[rounded corners, rectangle, draw]{$\delta$}
       }
;
\end{tikzpicture}
 \end{minipage}

Technically, the rule above deletes node $\alpha$ and the subtree
rooted at $a$, and rename $\gamma$ by $\delta$.


Finally, the Tree-LTL property to be checked on
$(\Rr',\Delta^\prime)$ and initial tree $T_0$ is $\f={\bf
  G}(P_1 \rightarrow{\bf  F} \, P_2)$, where $P_1=[\text{root}]([\gamma])$ and $P_2=[\text{root}]([\delta])$. It is easy to see that  $T_1
\arr{*} T_2$ via $\Rr$ iff $\Rr'$ does not
satisfy $\f$ from the  initial tree $T_0$: TP $P_1$ can be generated
only if from $T_1$ we can
reach a tree via $\Rr$, that contains $T_2$ as prefix. But then, TP
$P_2$ cannot be generated only if from $T_1$ we reach \emph{exactly}
$T_2$ (cf.~second rule).
\qed

\end{document}